# La interstitial defect-induced insulator-metal transition in oxide heterostructures LaAlO$_3$/SrTiO$_3$

Jun Zhou[1, 2] Ming Yang[3, 4] Yuan Ping Feng[1, 4, *] and Andrivo Rusydi[1,2,5,†]

[1]*NUSNNI-NanoCore, Department of Physics, National University of Singapore, Singapore 117411, Singapore.*

[2]*Singapore Synchrotron Light Source, National University of Singapore, Singapore 117603, Singapore.*

[3]*Institute of Materials Research and Engineering, A\*-STAR, 2 Fusionopolis Way, Singapore 138634, Singapore.*

[4]*Centre for Advanced 2D Materials and Graphene Research Centre, National University of Singapore, Singapore 117546.*

[5]*National University of Singapore Graduate School for Integrative Sciences and Engineering (NGS), 28 Medical Drive, Singapore 117456, Singapore*

[†,§] Correspondence should be addressed to: A.R. (phyandri@nus.edu.sg) or Y.P.F (phyfyp@nus.edu.sg)

Perovskite oxide interfaces have attracted tremendous research interest for their fundamental physics and promising all-oxide electronics applications. Here, based on first-principles calculations, we propose a novel surface La interstitial promoted interface insulator-metal transition in $LaAlO_3/SrTiO_3$ (110). Compared with surface oxygen vacancies, which play a determining role on the insulator-metal transition of $LaAlO_3/SrTiO_3$ (001) interfaces, we find that surface La interstitials can be more experimentally realistic and accessible for manipulation and more stable in ambient atmospheric environment. Interestingly, these surface La interstitials also induce significant spin-splitting states with Ti $d_{yz}/d_{xz}$ character at conducting $LaAlO_3/SrTiO_3$ (110) interface. On the other hand, for insulating $LaAlO_3/SrTiO_3$ (110) (<4 unit cells $LaAlO_3$ thickness), a distortion between La (Al) and O atoms is found at the $LaAlO_3$ side, partially compensating the polarization divergence. Our results reveal the origin of metal-insulator transition in $LaAlO_3/SrTiO_3$ (110) heterostructures, and also shed light on the manipulation of the superior properties of $LaAlO_3/SrTiO_3$ (110) for new possibilities of electronic and magnetic applications.

*Introduction.* Metal-insulator transition is one of the most important phenomena in broad communities for its fundamental research interest and diverse applications as it can be controlled *via* temperature, strain, external magnetic field and doping level [1]. Typically, metal-insulator transition occurs as a bulk property with a huge change of conductivity in correlated electron systems such as manganites [2] and vanadium dioxide [3]. About a decade ago, a different type of metal-insulator transition with two-dimensional electron gas (2DEG) was observed at the interface of perovskite oxide heterostructures, attracting tremendous research interest [4-9]. A paradigm system is the interface of two wide gap insulators, $LaAlO_3$ on $SrTiO_3$ ($LaAlO_3/SrTiO_3$) in (001) orientation [7]. Compared with the bulk metal-insulator transition, this interface conductivity has been reported to own several advantages. First, with a separation of doping resource (on $LaAlO_3$ side) and the free carriers (on $SrTiO_3$ side), the interface electron mobility is high [8]. Second, the interface conductivity could be manipulated by the film thickness, external electric field [7] and integrating ferroelectric materials [10,11]. Based on this behaviour, a device concept has been proposed by 'writing' and 'erasing' nanowires at the $LaAlO_3/SrTiO_3$ interface with the tip of a conducting atomic force microscope [12,13], providing a possible route for devices with ultrahigh-density. Third, it allows remote modifications on surface, like protonation [14], adsorbates [15,16], and metal layers [17,18], which significantly change the behavior of interface insulator-metal transition.

This remarkable $LaAlO_3$-thickness ($d_{LAO}$) dependent insulator-metal transition of $LaAlO_3/SrTiO_3$ (001) has been found to be determined by the formation of surface oxygen vacancies ($V_O$) of $LaAlO_3$ [19-23]. Such $V_O$ compensate the polarization divergence of $LaAlO_3/SrTiO_3$ (001) and induce a charge transfer into the interface, yielding a conducting interface [24]. However, for devices and applications, the manipulation and, particularly, characterization of $V_O$ in LaAlO$_3$/SrTiO$_3$ remains experimentally challenging due to the small atomic radius of oxygen, and the tuned interface conductivity suffers from instability problems. For example, the conducting nanowires at the interface of LaAlO$_3$/SrTiO$_3$ (001) produced by the conducting atomic force microscope are only stable for around 24 hours at room temperature in the air [12,25]. Therefore, for a more sustained manipulation, an interface insulator-metal transition involving more inert defects to ambient atmospheric environment is needed for practical electronics applications.

Recent experiments have demonstrated an unexpected insulator-metal transition in LaAlO$_3$/SrTiO$_3$ along (110) orientation [26-28]. Unlike the *step-like* insulator-metal transition of LaAlO$_3$/SrTiO$_3$ (001) [7], there are intermediate states around 4 unit cells (uc) of LaAlO$_3$ before the conductivity reaches relatively constant high value in thicker LaAlO$_3$. This indicates a more complicated mechanism governs the distinctive insulator-metal transition of LaAlO$_3$/SrTiO$_3$ (110). Furthermore, the 2DEG at LaAlO$_3$/SrTiO$_3$ (110) interface is totally different from LaAlO$_3$/SrTiO$_3$ (001) with unique Rashba spin-orbit fields [29,30], giant crystalline anisotropic magnetoresistance [27,31], anisotropy [27], orbital hierarchies and distribution [29,32]. To understand this distinctive insulator-metal transition of LaAlO$_3$/SrTiO$_3$ (110) and to explore new possibilities at this less studied interface, a comprehensive study is essential.

Here, *via* first-principles calculations, we propose a novel surface La interstitial ($I_{La}$) promoted insulator-metal transition and magnetism in LaAlO$_3$/SrTiO$_3$ (110), which possesses superior properties for practical electronic and spintronics applications than the surface $V_O$ determining systems (See the Supplemental Material [33] for details and justification of the calculation methodology based on VASP [22,27,34-38]). In order to justify our proposed model, we compare our theoretical study with existing experimental data and go beyond. Intriguingly, new surface spin-splitting hole states, interface magnetism and polarization distortions between La (Al) and O atoms are also shown.

*Stoichiometric $LaAlO_3/SrTiO_3$ (110).* We first perform calculations on the stoichiometric $LaAlO_3/SrTiO_3$ (110) (without any defect) with 2, 3, 4, 5, 6, 7, and 8 uc of $LaAlO_3$ respectively. As shown in Fig. 1(b), an insulator-metal transition is reproduced with a critical thickness of 5 uc $LaAlO_3$. This can be explained by the compensation of the polarization potential divergence caused by the polar discontinuity of this system. It is noted that $LaAlO_3/SrTiO_3$ (110) heterostructures with planar interfaces do not have polar discontinuity. However, buckled interfaces, as applied here, are more energetically favourable for $LaAlO_3/SrTiO_3$ (110) [27,39]. Experimental evidence also demonstrated the coexistence of La and Ti at the interface of $LaAlO_3/SrTiO_3$ (001) by high-angle annular dark-field scanning transmission electron microscopy (HAADF-STEM) and EELS measurements [26]. With buckled interfaces, the polar discontinuity arises in $LaAlO_3/SrTiO_3$ (110) (see Fig. S1 and the related discussions in Supplementary Material [33]). The polar discontinuity leads to an electric potential in $LaAlO_3$ which increases as $d_{LAO}$ increases. When the polarization potential exceeds the band gap of $LaAlO_3$, a Zener breakdown happens with a charge transfer from $LaAlO_3$ surface to $LaAlO_3/SrTiO_3$ (110) interface. The interface charge density further increases monotonically with $d_{LAO}$ to compensate the increased electric potential [see Fig. 1(b)]. In previous calculations with a stoichiometric model [27], the critical thickness of $LaAlO_3$ was claimed to be 4 uc but the authors emphasized that the number may vary with the calculation details. Here, 4 uc $LaAlO_3/SrTiO_3$ is still insulating with a very small energy gap (0.03 eV).

Interestingly, the charge transfer between LaAlO$_3$ and SrTiO$_3$ in thicker LaAlO$_3$/SrTiO$_3$ leads to excess electrons at interface and simultaneously excess holes at surface. The projected density of states (PDOSs) of surface oxygen atoms in 5 uc LaAlO$_3$/SrTiO$_3$ (110) is shown in Fig. 1(c). O1 indicate the oxygen atoms in the top AlO sub-layer, which are bonded with the Al atoms in the [001] direction [Fig. 1(a)] The uncompensated $p_x$ and $p_y$ orbitals contribute to the hole states [Fig. 1(c)]. The oxygen atoms O2 and O3 in the top O$_2$ and LaAlO sub-layers [Figs. 1(a) and (c)] are also hole-conducting but with decreased charge densities, which are contributed by $p_z$ orbitals for O2, and $p_x$/$p_y$ orbitals for O3, respectively. These hole states have the same total amount of charge density as the interface electrons, keeping the whole system neutral. Remarkably, these surface holes are spin polarized around Fermi level with a half-metal manner. Thus, for the stoichiometric LaAlO$_3$/SrTiO$_3$ (110) at this $d_{LAO}$, new surface magnetic ordering can be realized.

However, surface holes and monotonically increased interface charge density have not been observed experimentally [26,27]. This suggests that the stoichiometric model is not sufficient to explain the existing experimental results [26,27]. Different from the surface holes of the LaAlO$_3$/SrTiO$_3$ (001) heterostructure, which locate only on the oxygen atoms in the flat surface of the AlO$_2$ sub-layers, the unique surface hole distribution in the buckled LaAlO$_3$/SrTiO$_3$ (110) heterostructure opens up possibilities for the formation of surface interstitials, in particular, La interstitials (A site of ABO$_3$ perovskite oxide) [Fig. 3(a)]. Such La interstitials can directly bond with O1, O2, and O3 and compensate these holes.

*Surface interstitials.* Polar-induced defects have been introduced to the oxide heterostructures with polar discontinuity, successfully explaining the experimentally observed electronic and magnetic phenomena [19,20,22]. Here, we discuss the most possible mechanism among the polar-induced surface oxygen vacancies ($V_O$), La interstitials ($I_{La}$), and Al interstitials ($I_{Al}$) in LaAlO$_3$/SrTiO$_3$ (110). The formation energies ($E_f$, which is calculated with a convergence criterion of 0.02 eV) of $V_O$ and $I_{La}$ for 2 uc LaAlO$_3$/SrTiO$_3$ (110) are comparable while that of $I_{Al}$ is around 3 eV higher, implying a less favourable presence of surface $I_{Al}$. Thus we do not further consider $I_{Al}$ for the thicker LaAlO$_3$ on SrTiO$_3$ (110). Figure 2 shows the formation energies of surface $I_{La}$ and $V_O$ in 2, 3, 4, 5 and 6 uc LaAlO$_3$ on SrTiO$_3$ (110) in sample growth conditions [27]. The $E_f$ of $I_{La}$ and $V_O$ both decrease with $d_{LAO}$ and reach a critical value of around 0 eV at 4 uc LaAlO$_3$ for $I_{La}$ and 5 uc for $V_O$. The $E_f$ of $I_{La}$ and $V_O$ are comparable at 2 uc $d_{LAO}$, but the $E_f$ of $I_{La}$ become increasingly lower than that of $V_O$ for thicker LaAlO$_3$.

The lower $E_f$ of surface $I_{La}$ than surface $V_O$ can be understood as follows. The polar electric field in LaAlO$_3$ pushes the electrons produced by surface defects into the LaAlO$_3$/SrTiO$_3$ (110) interface, leaving the surface positively charged. These positively charged surface and excess interface electrons fully compensate the polar potential divergence of LaAlO$_3$/SrTiO$_3$ (110). At 5 uc LaAlO$_3$/SrTiO$_3$ (110), the $E_f$ of surface $I_{La}$ is around 0.69 eV lower than surface $V_O$, taken the same sample growth condition; and this energy difference further increases with $d_{LAO}$. Thus, in thick LaAlO$_3$/SrTiO$_3$, the surface $I_{La}$ (with much lower $E_f$) are more likely to form than surface $V_O$. Besides, surface $I_{La}$ excludes the possibility of further surface $V_O$ formation. The $E_f$ of surface $V_O$ would be as high as 2.33 (2.53) eV in oxygen poor (rich) condition when the polar potential divergence of 5 uc LaAlO$_3$/SrTiO$_3$ (110) has already been compensated by surface $I_{La}$. These results strongly suggest surface $I_{La}$ rather than surface $V_O$ dominate in the conducting LaAlO$_3$/SrTiO$_3$ (110).

Experimentally, the insulator-metal transition of LaAlO$_3$/SrTiO$_3$ (110) occurs at around 4uc LaAlO$_3$/SrTiO$_3$ (110) which has a sheet conductivity of one order of magnitude smaller than that of thicker LaAlO$_3$/SrTiO$_3$ (110) [26,27]. The $E_f$ of $I_{La}$ with a concentration of one third per uc is positive but small [0.26 eV (0.56 eV) in La rich (poor) conditions] at this thickness. These $E_f$ values imply a possible presence of $I_{La}$ at a lower concentration with a lower interface conductivity as observed experimentally [26,27]. From these analyses, we propose a model based on surface $I_{La}$ to explain the conductivity in LaAlO$_3$/SrTiO$_3$ (110).

*LaAlO$_3$/SrTiO$_3$ (110) with surface $I_{La}$.* The electronic properties of 5 uc LaAlO$_3$/SrTiO$_3$ (110) are shown in Fig. 3. The in-plane averaged partial charge density in Fig. 3(b) shows that the excess electrons locate at the SrTiO$_3$ side of the heterostructure with a maximum distribution on the interface LaTiO sub-layer. This charge density decreases to the inner SrTiO sub-layers but increases again from the fourth to the middle SrTiO sub-layers in the SrTiO$_3$ (110) substrate. This hump of charge density in the centre layers of SrTiO$_3$ is often found in the LaAlO$_3$/SrTiO$_3$ (001) heterostructures by GGA approximations [40,41], which does not affect our main conclusions here. The total amount of interface charge density of one electron per uc, is required to a full compensation of the polar potential divergence in LaAlO$_3$/SrTiO$_3$ (110). Besides, the surface becomes insulating now [Fig. 3(b)], consistent with reported experimental observations [26,27].

The PDOS in Fig. 3(c) shows the excess electrons on the interface Ti are mainly contributed by $d_{xz}$ and $d_{yz}$ orbitals, different from LaAlO$_3$/SrTiO$_3$ (001) interfaces [42,43]. This orbital hierarchy was observed in recent experimental results [32], and may affect the residual carrier density deduced from transport measurements. Considering the $\sqrt{2}$ times area of SrTiO$_3$ (110) unit cell as that of SrTiO$_3$ (001), the carrier density in LaAlO$_3$/SrTiO$_3$ (110) (one electron per uc) should be $\sqrt{2}$ times as large as that in LaAlO$_3$/SrTiO$_3$ (001) heterostructures (half electrons per uc). However, experimental results have shown that LaAlO$_3$/SrTiO$_3$ (110) samples have slightly decreased residual carrier densities than LaAlO$_3$/SrTiO$_3$ (001) samples [26,27]. The residual carrier density of LaAlO$_3$/SrTiO$_3$ (001) from transport measurements is much lower than half electrons per uc as predicted theoretically [19,20,22,24] and measured by high-energy optical conductivity [44]. This was explained by the localized electrons in this interface [43,45]. The reduced residual carrier densities of the LaAlO$_3$/SrTiO$_3$ (110) samples imply even less electrons contribute to the carrier density in this interface. This is reasonable as the dominant $d_{xz}$ and $d_{yz}$ orbitals are more localized than the $d_{xy}$ [45].

*Interface magnetism.* The high interface charge density of LaAlO$_3$/SrTiO$_3$ (110) from the strong correlated orbitals of Ti atoms leads to possible interface magnetism. Our calculations predict a total magnetic moment of ~0.56 $\mu_B$ per unit cell at 5 uc LaAlO$_3$/SrTiO$_3$ (110) with surface $I_{La}$. This significant magnetic moment would induce stronger experimentally observable magnetic signals than that of LaAlO$_3$/SrTiO$_3$ (001) [46]. The PDOS shown in Fig. 3(c) suggests spin-splitting states on Ti atoms of LaAlO$_3$/SrTiO$_3$ (110) interfaces. Contrary to the main $d_{xy}$ contribution of magnetism at LaAlO$_3$/SrTiO$_3$ (001) [46], the magnetic moments here are mainly from the Ti degenerate $d_{xz}$ and $d_{yz}$ orbitals.

*Polar distortion.* Polar distortion between cationic and anionic atoms is another way to compensate the polar potential divergence of oxide heterostructures [47-49], but have not been reported in LaAlO$_3$/SrTiO$_3$ (110). Our calculations show that polar distortions between Al (La) and oxygen atoms also exist in insulating LaAlO$_3$/SrTiO$_3$ (110) (without any defect). Generally speaking, the oxygen, La and Al atoms in LaAlO$_3$ move towards LaAlO$_3$ surface due to the relaxation of thin LaAlO$_3$ film on SrTiO$_3$ (110) substrate. However, a clear pattern of polar distortions between anions (oxygen) and cations (La and Al) in LaAlO$_3$ is found [see Figure 4(a)]. The relative displacements between La/Al and oxygen atoms induce electric dipoles, which are in opposite directions (except the atoms in the top AlO sub-layer due to the surface effects) with the inner electric field produced by the polar discontinuity in LaAlO$_3$/SrTiO$_3$ (110) and thus partly compensate the polar potential divergence. These polar distortions, if observed experimentally in thin LaAlO$_3$/SrTiO$_3$ (110) films ($d_{LAO}$ < 4 uc), would be a strong support for the polar discontinuity and the buckled interface of LaAlO$_3$/SrTiO$_3$ (110). In contrast, on the SrTiO$_3$ (110) side, all the Sr, Ti and O atoms move a bit to the interface, but there is no relative displacement between anions and cations.

When surface $I_{La}$ are introduced to LaAlO$_3$/SrTiO$_3$ (110) (lower panel in Fig. 4), these polar distortions between La (Al) and O atoms disappear (except the atoms in the top AlO and LaAlO sub-layers due to the surface effects). This is a proof that the polar potential divergence of LaAlO$_3$ is completely compensated by the surface $I_{La}$ incuded charge transfer. On the SrTiO$_3$ (110) side, opposite displacements between Sr or Ti and O atoms with that of the La or Al and O atoms in LaAlO$_3$ of the stoichiometric LaAlO$_3$/SrTiO$_3$ (110) are shown. Furthermore, the experimentally observed large Rashba spin-splitting in LaAlO$_3$/SrTiO$_3$ (110) [30] can be induced by these large polar distortions between Ti and O in SrTiO$_3$ [30,50,51], which in turn supports the surface La interstitials over the stoichiometric picture as well.

*Discussion.* We have established a comprehensive picture to understand the distinctive insulator-metal transition in LaAlO$_3$/SrTiO$_3$ (110) based on polar distortions and surface $I_{La}$. Compared with oxygen vacancies, accurate manipulation of surface $I_{La}$ (heavy element) can be more experimentally realistic and accessible, because La cations are observable experimentally such as by HAADF-STEM [26]. This is a significant advantage for the manipulation as it provides direct experimental feedbacks. Besides, the concentration of surface $I_{La}$ can be controlled by changing La/Al ratio during sample growth. The primarily control of La/Al ratio in experiments has been demonstrated at LaAlO$_3$/SrTiO$_3$ (001) [52-54].

Moreover, unlike vacancies, the formation of La interstitials requires extra La sources. Once the manipulation with desired distribution of surface $I_{La}$ is completed, the existing surface $I_{La}$ are inert from environment change due to their negative formation energies and no new random surface $I_{La}$ will form at the insulating region due to the lack of source. Thus a manipulation through surface $I_{La}$ has superior properties than $V_O$ for practical electronic applications. Surface $I_{La}$ is a new degree of freedom to tune the electronic and magnetic properties of LaAlO$_3$/SrTiO$_3$ (110). For example, by accurate control of La sources during the sample growth process, stoichiometric LaAlO$_3$/SrTiO$_3$ (110) films may be realized at around 4 or 5 uc LaAlO$_3$, in which the $E_f$ of other defects like $V_O$ is still positive. With this stoichiometric LaAlO$_3$/SrTiO$_3$ (110), the new surface spin-splitting hole states [Fig. 1(c)] can be realized experimentally.

*Conclusion.* In summary, we have proposed a novel surface La interstitials promoted insulator-metal transition in LaAlO$_3$/SrTiO$_3$ (110). Compared with surface oxygen vacancies, surface interstitials may be more experimentally realistic and accessible for engineering during sample growth and more stable in ambient atmospheric environment, promising for precise tuning of the emergent properties at oxide heterostructures. Furthermore, our calculations show new spin-splitting hole states at the surface of stoichiometric LaAlO$_3$/SrTiO$_3$ (110), magnetism at the interface of LaAlO$_3$/SrTiO$_3$ (110) with surface $I_{La}$, and polar distortions between La (Al) and O atoms in insulating LaAlO$_3$/SrTiO$_3$ (110) interfaces. Our result opens new possibilities for manipulating properties of LaAlO$_3$/SrTiO$_3$ (110) and other perovskite oxide heterostructures by accurate control of robust surface interstitials.

This work is supported by the Singapore National Research Foundation under its Competitive Research Funding (No. NRF-CRP 8-2011-06 and No. R-398-000-087-281), MOE-AcRF Tier-2 (MOE2015-T2-1-099, MOE2015-T2-2-065, and MOE2015-T2-2-147), FRC (R-144-000-368-112), and 2015 Merlion Project. We thank Centre for Advanced 2D Materials and Graphene Research Centre at National University of Singapore to provide the computing resource.

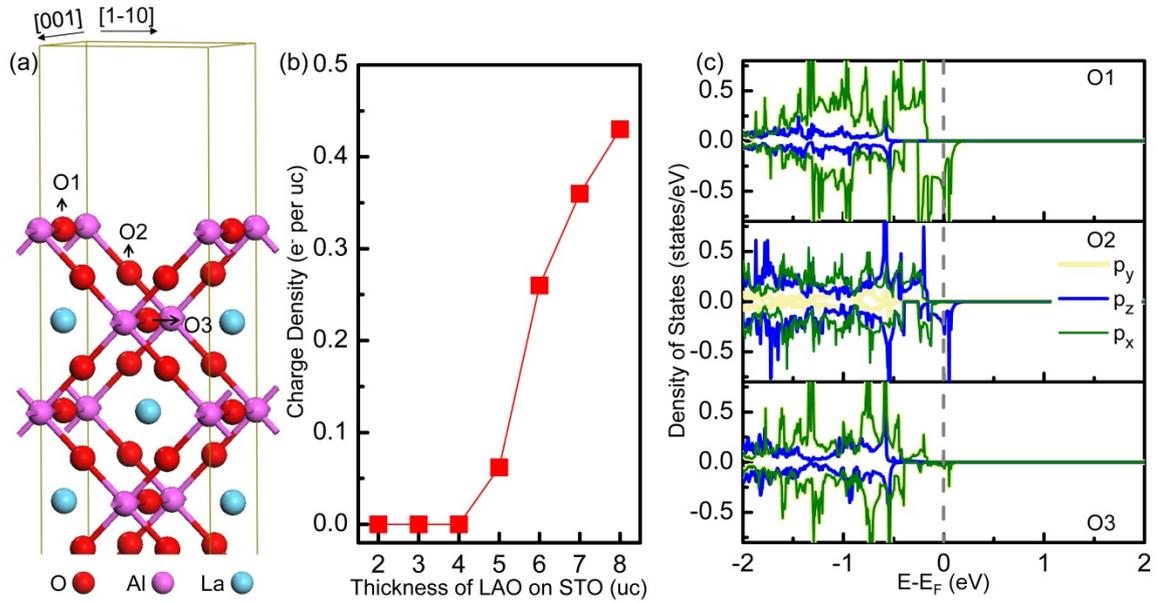

FIG. 1. Structural and electronic properties of stoichiometric LAO/STO (110). (a) Structural guidance for the stoichiometric LAO/STO (110) with buckled interface. Due to the stoichiometry, the LAO surfaces are also buckled. For clarity, only the top LAO layers are shown here. The oxygen atoms in the top AlO, O$_2$, and LaAlO sub-layers are marked with O1, O2, and O3, respectively. (b) Interface carrier density for stoichiometric 2, 3, 4, 5, 6, 7 and 8 uc LAO/STO (110). (c) Projected Density of States for O1, O2, and O3 as shown in Fig. 1(a).

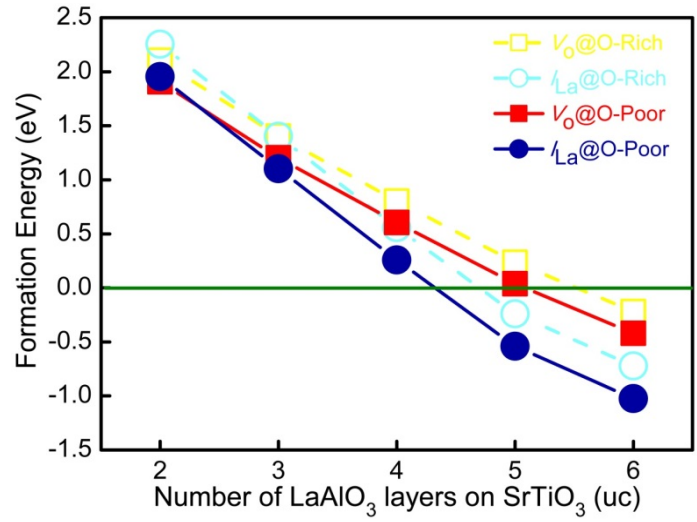

FIG. 2. Formation energy for 2, 3, 4, 5 and 6 uc LaAlO$_3$/SrTiO$_3$ (110) with surface oxygen vacancies and surface La interstitials respectively in both oxygen rich and oxygen poor conditions.

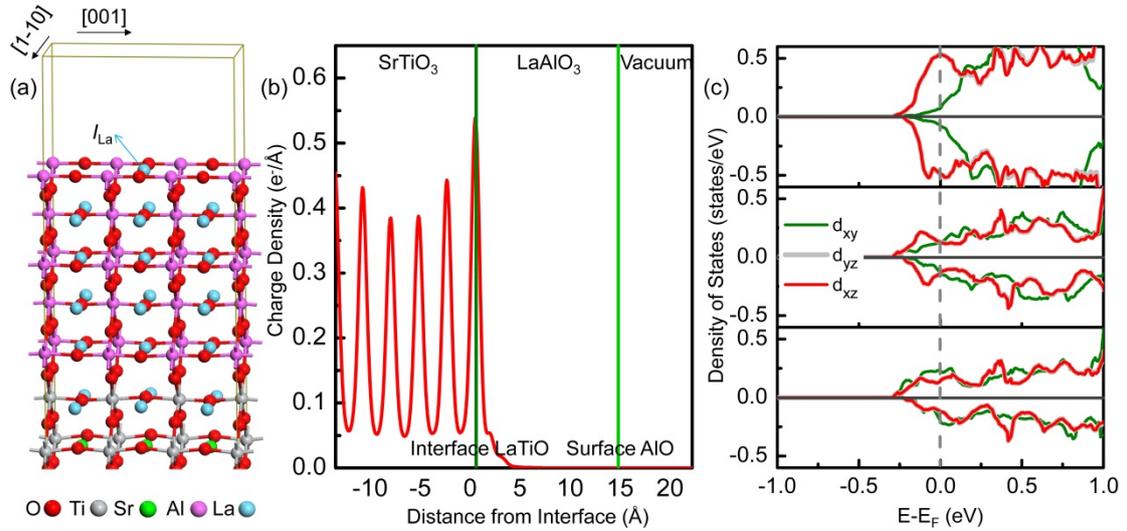

FIG. 3. Structural and electronic properties of LaAlO$_3$/SrTiO$_3$ (110) with surface La interstitial defects. (a) Structural guidance for 5 uc LaAlO$_3$/SrTiO$_3$ (110) with surface La interstitials. (b) In-plane averaged charge density along *c* axis for 5 uc LaAlO$_3$/SrTiO$_3$ (110) with surface La interstitials. The charge density is calculated by integrating the DOS of the conduction bands below the Fermi energy. (c) PDOS of Ti atoms in the interface (top panel), the third (middle panel) and the fifth (bottom panel) SrTiO layers of SrTiO$_3$ for 5 uc LaAlO$_3$/SrTiO$_3$ (110) with surface La interstitials.

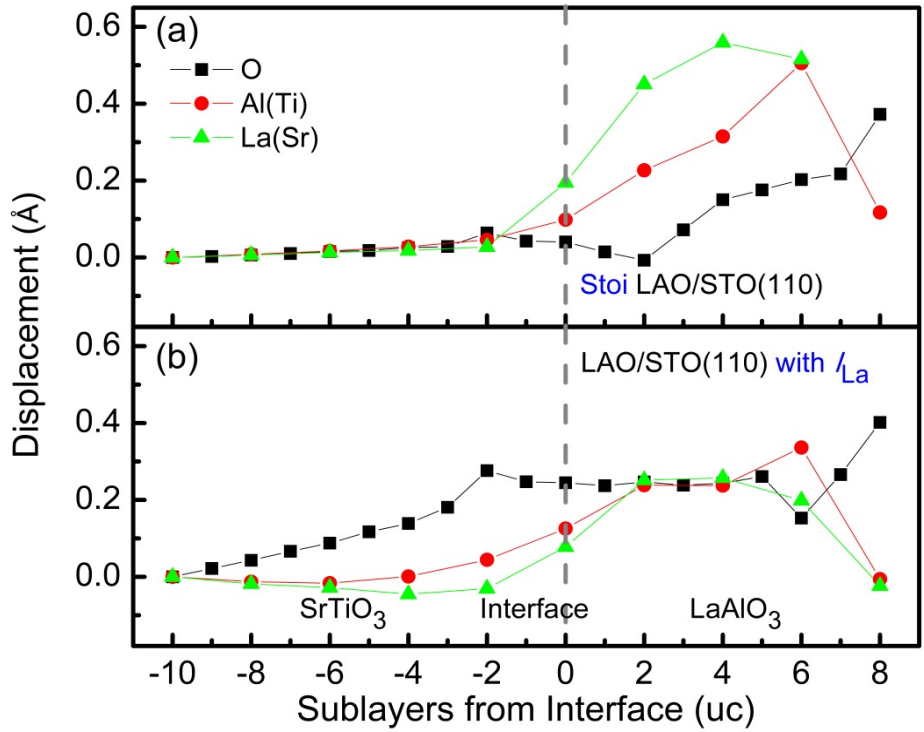

FIG. 4. Atomic displacements for LaAlO$_3$/SrTiO$_3$ (110) interfaces with respect to un-relaxed structures. (a). A profile of atomic displacements along $c$ axis for stoichiometric 4 uc LaAlO$_3$/SrTiO$_3$ (110). (b). A profile of atomic displacements along $c$ axis for 4 uc LaAlO$_3$/SrTiO$_3$ (110) with surface La interstitials. Displacements of O, Al (Ti) and La (Sr) are shown by black square, red circle and green triangle, respectively. 'Stoi' and 'with $I_{La}$' highlighted in blue indicate the stoichiometric LaAlO$_3$/SrTiO$_3$ and LaAlO$_3$/SrTiO$_3$ (110) with surface La interstitials, respectively.

Page **17** of **21**

Page **17** of **21**